\documentclass{article}

\usepackage{arxiv}
\usepackage{listings}
\usepackage[utf8]{inputenc} 
\usepackage[T1]{fontenc}    
\usepackage{hyperref}       
\usepackage{url}            
\usepackage{booktabs}       
\usepackage{amsfonts}       
\usepackage{nicefrac}       
\usepackage{microtype}      
\usepackage{lipsum}
\usepackage{graphicx}
\graphicspath{ {./images/} }

\title{Daany - DAta ANalYtics on .NET }

\author{
 Bahrudin Hrnjica \\
  Faculty of Engineering Sciences\\
  University of Bihac\\
  Bihac, 77000, Bosnia and Herzegovina,  \\
  \texttt{{name}.{surname}unbi.ba} \\
}

\begin{document}
\maketitle
\begin{abstract}
 \texttt{Daany} is .NET and cross platform data analytics and linear algebra library written in C\# supposed to be a tool for data preparation, feature engineering and other kind of data transformations and feature engineering. The library is implemented on top of .NET Standard 2.1 and supports .NET Core 3.0 and above separated on several Visual Studio projects that can be installed as a NuGet package.The library implements `DataFrame` as the core component with extensions of a set of data science and linear algebra features. The library contains several implementation of time series decomposition (SSA, STL ARIMA), optimization  methods (SGD) as well as plotting support. The library also implements set of features based on matrix, vectors and similar linear algebra operations. The main part of the library is the \texttt{Daany.DataFrame} with similar implementation that can be found in python based Pandas library. The paper presents the main functionalities and the implementation behind the \texttt{Daany} packages in the form of developer guide and can be used as manual in using the \texttt{Daany} in every-day work. To end this the paper shows the list of papers used the library. 
\end{abstract}


\section{Introduction \texttt{Daany}}

\texttt{Daany} is .NET  data analytic library written in C\# with support various kind of data transformation, descriptive statistics and linear algebra. With \texttt{Daany}  an user can load the data from text based file into the \texttt{DataFrame} arranged into columns, rows and index. The user can also create \texttt{Series} object - a special kind of \texttt{Daany.DataFrame} in order to work with time series data. Once the data is loaded the user can start analyzing the data by performing various transformation and results can be display as chart or tabular data. 

\texttt{Daany} is an open source project hosted at https://github.com/bhrnjica/daany. In order to build and run the library from the source code, one can use Windows or Linux based distribution. The most easiest way to build the library is to use command line dotnet tool and build the binaries. Otherwise the  \textbf{Visual Studio} and \textbf{Visual Studio Code} can be used for building and developing the library. The GitHub project contains documentation and unit test project for any implemented operation in the library. Also the \href{https://github.com/bhrnjica/daany/blob/master/docs/DevGuide/developer_guide.md}{\textbf{Daany Developer Guide}} contains in details all aspects of the implementation and library features. 
 
The library implements the \texttt{Daany.MathStuff} module which consists of of algebraic operations on matrix and vectors as well as rich set of statistics distributions and parameters. Furthermore \texttt{Daany.LinA} extends it in order to gain better performance and functionalities. The \texttt{Daany.LinA} is the .NET wrapper around the LAPACK \cite{intelLapack} and BLAS \cite{intelBlass} C++ libraries. Besides data analysis, the library implements a set of statistics or data science features e.g. time series decomposition, optimization performance parameters and similar. The main components of the library which can be installed separately as a NuGet package are:
 
 \begin{itemize}
     \item \texttt{Daany.DataFrame},
     \item \texttt{Daany.DataFrame.Ext}, 
     \item \texttt{Daany.Stat}, 
     \item \texttt{Daany.LinA}.
 \end{itemize}
 
\section{Daany Architecture}
\label{sec:headings}

\begin{figure}
  \centering
   \includegraphics{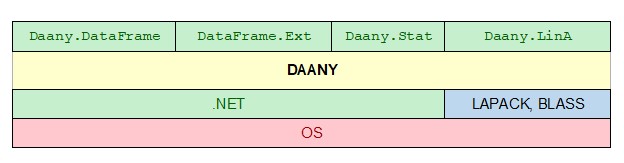}
  \caption{Architecture of Daany library}
  \label{fig:fig1}
\end{figure}

Daany is classic .NET component implemented through the several visual studio projects. The library is based on .NET Framework and Intel MKL implementation of LAPACK and BLASS libraries. The architecture diagram of the library is described at Figure \ref{fig:fig1}. 

Beside as the classic .NET library usage \texttt{Daany} is implemented to be used for data exploration and transformation with \texttt{.NET Jupyter Notebook}.   

\begin{figure}
  \centering
   \includegraphics[width=0.95\textwidth]{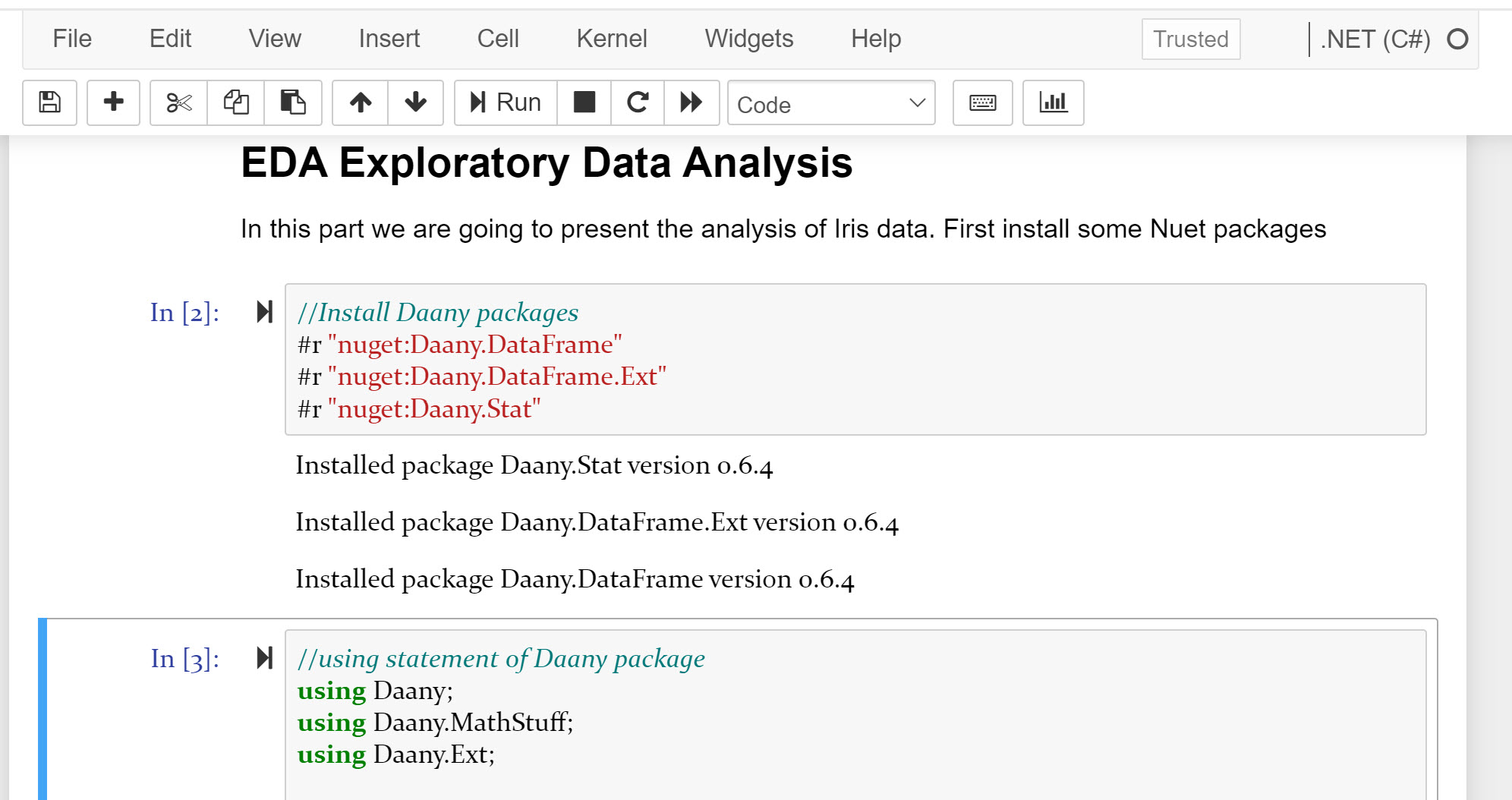}
  \caption{Using Daany library in .NET Jupyter Notebook}
  \label{fig:fig2}
\end{figure}

When using \texttt{Daany} in the Jupyter Noebook a user should register the formatter for the \texttt{DataFrame} in order the notebook can render the DataFrame as natural tabular data. The formatter can be found later in the text.

\subsection{How to start with \texttt{Daany}}

\texttt{Daany} is .NET component and can be run on any platform .NET  supports. It can be used by Visual Studio or Visual Studio Code. It is consisted of the five NuGet packages, so the easiest way to start with is to install the packages in the .NET application or Jupyter Notebook. 

\begin{figure}
  \centering
   \includegraphics[width=0.95\textwidth]{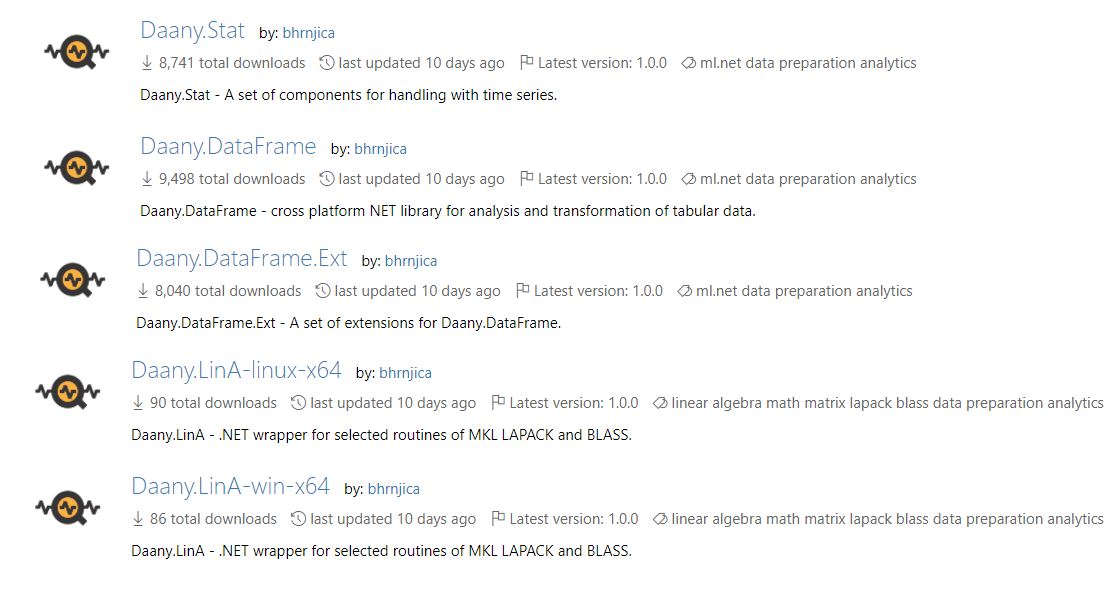}
  \caption{Nuget packages of Daany library}
  \label{fig:fig3}
\end{figure}

In the Nuget repositories the user can find five packages starting with Daany, which are listed at Figure \ref{fig:fig3}.

\begin{enumerate}
     \item \texttt{Daany.DataFrame} - one can use this option only for data analysis and data transformation by using data frame,
     \item \texttt{Daany.DataFrame.Ext} - the package sontains set of DataFrame extensions for \texttt{DataFrame} in form of plotting support, column data transformations etc, 
     \item \texttt{Daany.Stat}  the package already contains \texttt{Daany.DataFrame}, as well as time series decomposition and related statistics features.
     \item \texttt{Daany.LinA} - the package is .NET wrapper round LAPACK and BLASS routines from the MKL. The package has two versions: 
     \begin{enumerate}
        \item \texttt{Daany.LinA-win-64} - windows 64 version,
        \item \texttt{Daany.LineA-linux-64} - Linux based distributions.
    \end{enumerate}
 \end{enumerate}
 
Once you install the packages, you can start developing your app using Daany packages.

\subsection{Namespaces in \texttt{Daany}}

\texttt{Daany} project contains several namespaces for separating different implementation. The following list contains relevant namespaces:

\begin{enumerate}
    \item  \texttt{using Daany} – data frame and related code implementation,
    \item  \texttt{using Daany.Ext} – data frame extensions, used with dependency on third party libraries,
    \item  \texttt{using Daany.Stat} – set of statistics related implementations e.g. descriptive statistice, optimizers, time series etc.
     \item  \texttt{`using Daany.LineA` } – set of Linear algebra routines for using LAPACK and BLASS.
 \end{enumerate}

\section{\texttt{Daany.DataFrame}  - data analysis and transformation of tabular data }

\begin{figure}
  \centering
   \includegraphics[width=0.95\textwidth]{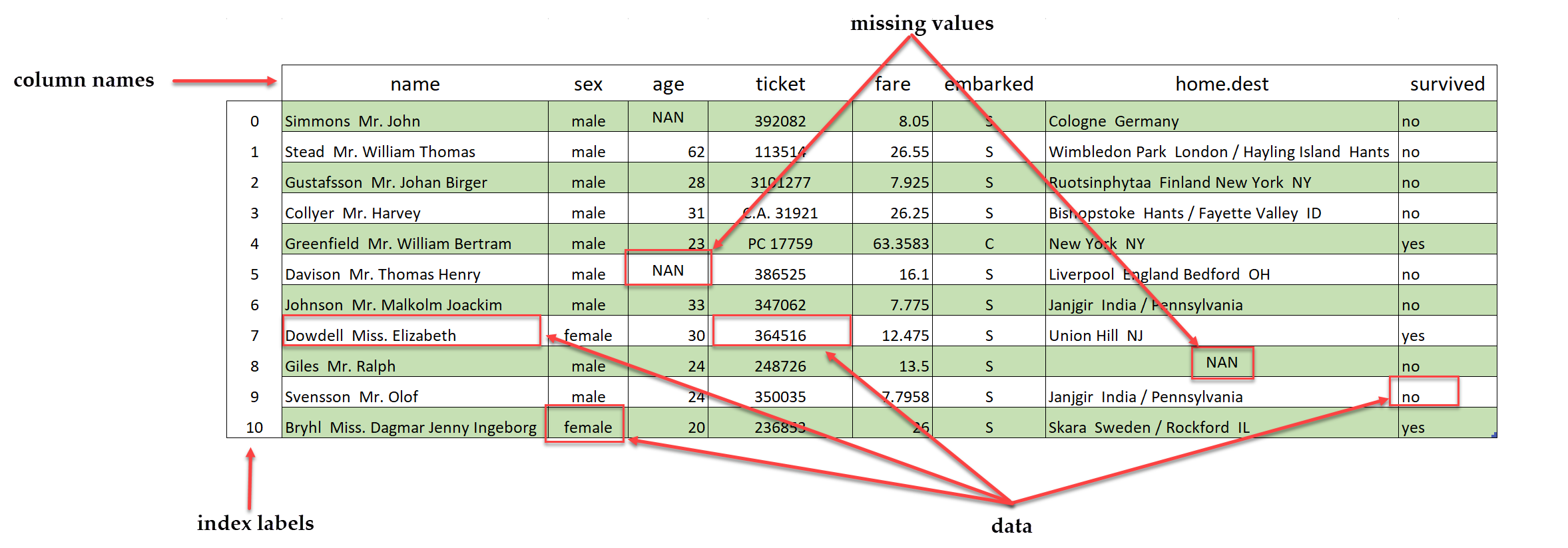}
  \caption{\texttt{DataFrame} components }
  \label{fig:fig4}
\end{figure}

The main part of \texttt{Daany} project is \texttt{Daany.DataFrame} -  an \texttt{c\#} implementation of data frame. A data frame is software component used for handling tabular data, especially for data preparation, feature engineering and analysis during development of machine learning models. The concept of \texttt{Daany.DataFrame} implementation is based on simplicity and .NET coding standard. It represents tabular data consisting of columns and rows. Each column has name and type and each row has its index and label. Usually, rows indicate a zero axis, while columns indicate axis one. Figure \ref{fig:fig4} shows a data frame structure

The basic components of the data frame are:

 \begin{enumerate}
   \item   \texttt{header} - list of column names,
   \item   \texttt{index}  – list of object representing each row,
   \item   \texttt{data} – list of values in the data frame,
   \item   \texttt{missing value} – data with no values in data frame.
 \end{enumerate}
 
In order to create a DataFrame there are several options:

 \begin{enumerate}
   \item from a list of values, by specifying column names and row count
    \item from a dictionary, letting keys be column names and values be column values,
    \item from text-based file, where each line represents row values,
    \item  as a return object for almost any data frame operations.
  \end{enumerate}

\subsection{Create \texttt{Daany.DataFrame} by loading data from a file}

By using static method \texttt{DataFrame.FromCsv} a user can create data frame object from the \texttt{csv} file. Otherwise, data frame can be persisted on disk by calling static method \texttt{DataFrame.ToCsv}. 

The following code shows how to use static methods \texttt{ToCsv} and \texttt{FromCsv} to show persisting and loading data to data frame:

\begin{lstlisting}

string filename = "df_file.txt";
//define a dictionary of data
var dict = new Dictionary<string, List<object>>
{
    { "ID",new List<object>() { 1,2,3} },
    { "City",new List<object>() { "Sarajevo", "Seattle", "Berlin" } },
    { "Zip Code",new List<object>() { 71000,98101,10115 } },
    { "State",new List<object>() {"BiH","USA","GER" } },
    { "IsHome",new List<object>() { true, false, false} },
    { "Values",new List<object>() { 3.14, 3.21, 4.55 } },
    { "Date",new List<object>() { DateTime.Now.AddDays(-20) , DateTime.Now.AddDays(-10) , DateTime.Now.AddDays(-5) } },

};

//create data frame with 3 rows and 7 columns
var df = new DataFrame(dict);

//first Save data frame on disk and load it
DataFrame.ToCsv(filename, df);

//create data frame with 3 rows and 7 columns
var dfFromFile = DataFrame.FromCsv(filename, sep:',');

//check the size of the data frame
Assert.Equal(3, dfFromFile.RowCount());
Assert.Equal(new string[] { "ID", "City", "Zip Code", "State", "IsHome", "Values", "Date" }, dfFromFile.Columns);
Assert.Equal(7, dfFromFile.ColCount());

\end{lstlisting}

First, the data frame is created from the dictionary collection. Then data frame is stored to file. After successfully saving, the same data frame is created the csv file. The end of the code snippet, several asserts are defined in order to prove everything is correctly implemented.

In case the performance is important, the column types should be pass to \texttt{FromCSV} method in order to achieve up to 50\% of loading time. 

For example the following code loads the data from the file, by passing predefined column types:

\begin{lstlisting}
//defined types of the column 
var colTypes1 = new ColType[] { ColType.I32, ColType.IN, ColType.I32, ColType.STR, ColType.I2, ColType.F32, ColType.DT };
//create data frame with 3 rows and 7 columns
var dfFromFile = DataFrame.FromCsv(filename, sep: ',', colTypes: colTypes1);
\end{lstlisting}

\subsection{Create \texttt{Daany.DataFrame} from a web url}

Data can be loaded directly from the web storage by using \texttt{FromWeb} static method. The following code shows how to load the \texttt{Concrete Slump Test} data from the web. The data set includes 103 data points. There are 7 input variables, and 3 output variables in the data set: \texttt{Cement}, \texttt{Slag}, \texttt{Fly ash}, \texttt{Water}, \texttt{SP}, \texttt{Coarse Aggr.},\texttt{Fine Aggr.}, \texttt{SLUMP (cm)}, \texttt{FLOW (cm)}, \texttt{Strength (Mpa)}. 

The following code load the \texttt{Concrete Slump Test} data set into \texttt{Daany.DataFrame}:

\begin{lstlisting}
//define web url where the data is stored
var url = "https://archive.ics.uci.edu/ml/machine-learning-databases/concrete/slump/slump_test.data";
//
var df = DataFrame.FromWeb(url);
df.Head(5)
\end{lstlisting}

Besides presented examples, the data frame can be created as a results of any operations applied on the Daany data frame.

\section{\texttt{Daany.DataFrame}  operations}

\texttt{Daany.DataFrame} has rich set of operations that can be classified into several groups:

\begin{enumerate}
    \item \textbf{Add or Insert new row/column in the DataFrame}, - this group of operations are used to create a new row or column in the existing DataFrame. A created column/row can be defined as list or dictionary collections, but also can be defined dynamically based on the calculation logic from the existing columns in the data frame.
    \item \textbf{Handling missing values in DataFrame} - this set of operations handle the missing values in the data frame. 
    \item \textbf{Aggregate} - this group of operations include performing arithmetic operation on data frame. The result of the aggregation is new list of values or new data frame containing the result of aggregation operations. 
    \item \textbf{Filter} -  operations return data frame with specific filter condition. Beside filter the \textbf{RemoveRows} - method acts opposite  and removes all rows with specified condition by using delegate implementation. 
    \item \textbf{Sorting} - used for sorting the rows in the Data frame. The sorting operation supports both ascending and descending order. 
    \item \textbf{Grouping} - used for grouping data in the data frame using one, two or three data frame's columns. Usually while grouping the data one can perform some rolling or aggregate operations which is fully supported.
    \item \textbf{Joining and merging the data frames} - this set of operations are used when one data frame needs to be created from two or more data frames. 
    \item \textbf{Selection in data frame} this set of operation are used when selecting one or more rows/columns/cells. 

\end{enumerate}

\section{\texttt{Daany.Stat} - Daany statistics}

The second main daany component is \texttt{Daany.Stat} contains implementation of descriptive statistics, opetimization, time series decomposition etc. 

The following list contains some of \texttt{Daany.Stat} features:

\begin{enumerate}

\item Conversion time series into data frame and Series
\item \textbf{Seasonal and Trend decomposition using Loess} - STL time series decomposition,
\item \textbf{Singular Spectrum Analysis} SSA time series decomposition,
\item Set of Time Series operations like moving average, etc....

\end{enumerate}

\bibliographystyle{unsrt}  


\end{document}